# Supporting Finite Element Analysis with a Relational Database Backend
# Part II: Database Design and Access

Gerd Heber
Cornell Theory Center, 638 Rhodes Hall,
Cornell University, Ithaca, NY 14853, USA
heber@tc.cornell.edu

Jim Gray
Microsoft Research, San Francisco, CA 94105, USA
Gray@microsoft.com

# Supporting Finite Element Analysis with a Relational Database Backend

## Part II: Database Design and Access

Gerd Heber*, Jim Gray†

*Cornell Theory Center, 638 Rhodes Hall, Ithaca, NY 14853, USA

heber@tc.cornell.edu

†Microsoft Research, 455 Market St., San Francisco, CA 94105, USA

Gray@microsoft.com

**Abstract**: This is Part II of a three article series on using databases for Finite Element Analysis (FEA). It discusses (1) db design, (2) data loading, (3) typical use cases during grid building, (4) typical use cases during simulation (get and put), (5) typical use cases during analysis (also done in Part III) and some performance measures of these cases. It argues that using a database is simpler to implement than custom data schemas, has better performance because it can use data parallelism, and better supports FEA modularity and tool evolution because database schema evolution, data independence, and self-defining data.

## Introduction

Computational materials science simulations are data-intensive tasks. They start by designing one or more geometric models of parts or structures and generating the simulation finite element mesh(es) or other discretizations. Then the simulation is run, and finally one analyzes the results. A large model can take days to design, weeks or months to run, and months to analyze. Each of these steps may have many subtasks and some of them might be performed in loops. Each requires careful bookkeeping, and each consumes and produces large and complex datasets.[1]

Traditionally, materials science simulations, and indeed most scientific tasks have been done using files as the basic data interchange metaphor. Each step of the workflow consumes and produces files – uninterpreted streams of bytes. Consequently, analysis tools have large portions of non problem-oriented code that do nothing more than parsing and reformatting byte-streams.

We believe basing workflows on uninterpreted byte streams (files) is the wrong approach. Rather, one wants strongly-typed self-defining semantic information stores at each workflow step and one wants a tool that makes it easy to write such strongly-typed self-defining data streams. Indeed, the program step's inputs and outputs are often not streams at all; rather they are spatial, temporal, or conceptual data subsets. So, scientific simulation workflows need to be able to read and write arbitrary subsets or dynamic aggregations of the data store.

Databases provide exactly these capabilities. They store strongly-typed (explicitly modeled!) data, they allow arbitrary subsets to be inserted, and they manage very large data stores. They allow multiple organizations and allow easy reorganization. Databases also have other useful features like security, schema evolution (the ability to change data design), protection from hardware and other failures, concurrency control, data scrubbing tools, data mining tools, and non-procedural query tools.

We are not saying that files are going away – after all, databases are typically stored in files. Rather we are saying that scientists in general, and FEA practitioners in particular, should be working at a higher level of abstraction, they should be able to think in terms of object collections that can be accessed by spatial, temporal, or other attributes. Again, this is an evolution of the concept of a byte stream to the more general notion of an object store.

We begin this report with a brief recap of Part I of this three-part series. Subsequently, we introduce three computational geometry problems frequently encountered in FEA and show how to solve the most common one using a modern relational database, SQL Server 2005. This is followed by a performance discussion and some practical considerations.

---

[1] In this report, we ignore the enormous body of data and knowledge produced by experiments, captured by sensors, and published in the literature. Despite access limitations to parts of this information due to commercial and/or national defense value, modeling and charting just the publicly accessible "materials universe" remains a formidable task.

## The FEA Workflow and Database

The major steps in the FEA workflow include (see [19]):

1. Model/geometry generation (CAD etc.)
2. Meshing[2] (= decomposition into simple shapes and voxels)
3. Attribute definition and assignment (material properties, boundary conditions)
4. Equation formulation
5. Solution – running the model to produce simulated behavior over time
6. Post-processing (e.g., visualization, feature detection, life prediction)

As explained in [3], the unstructured finite element analysis (FEA) mesh is part of the simulation's metadata. Although the mesh itself makes up the bulk of the metadata, it is typically only less than 5% of the total data (which include simulation results); yet, managing this relatively small subset is crucial for post-processing, analysis, and visualization. It is essential to have a flexible format for it, essential to be able to evolve it, and essential to be able to access it from many different views.

There is no standardized representation for finite element meshes: industrial FEA packages use proprietary file formats and end-users spend a good deal of time writing conversion routines and keeping their converters up-to-date with format changes. These file formats are typically optimized for storage space-efficiency and are very different from the layout of in-core structures, which are optimized for fast lookup and access. Designing a database schema for a mesh is much like creating the easy-to-access in-memory structures, with the exception that one does not think in terms of pointers, arrays, or objects, but rather thinks in terms of relations, attributes, and indices. In other words, file formats are *not* a good guide for schema design.

Figure 1 shows the core tables dealing with tetrahedral elements; for simplicity it focuses on the volume mesh relations (it ignores the surface mesh). As mentioned in Part I [3], there is more to an FEA model than just the mesh. There is typically some form of parametric geometry or a polyhedral geometric structure; for example, the polycrystal models described in Part III [4].

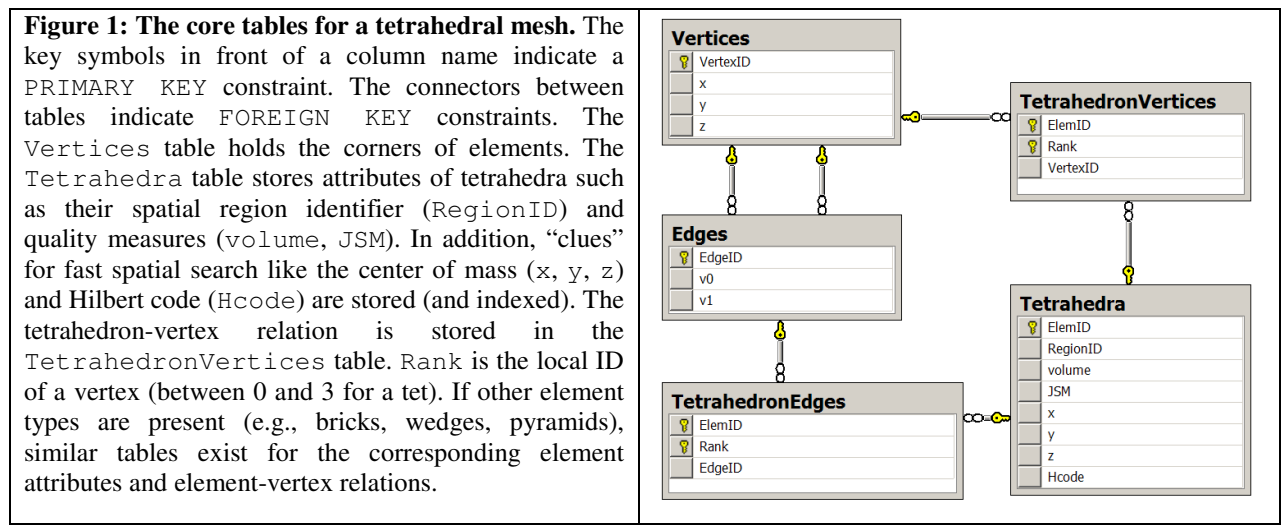

**Figure 1: The core tables for a tetrahedral mesh.** The key symbols in front of a column name indicate a `PRIMARY KEY` constraint. The connectors between tables indicate `FOREIGN KEY` constraints. The `Vertices` table holds the corners of elements. The `Tetrahedra` table stores attributes of tetrahedra such as their spatial region identifier (`RegionID`) and quality measures (`volume`, `JSM`). In addition, "clues" for fast spatial search like the center of mass (`x, y, z`) and Hilbert code (`Hcode`) are stored (and indexed). The tetrahedron-vertex relation is stored in the `TetrahedronVertices` table. `Rank` is the local ID of a vertex (between 0 and 3 for a tet). If other element types are present (e.g., bricks, wedges, pyramids), similar tables exist for the corresponding element attributes and element-vertex relations.

In Part I, the section "Mapping an FEA Data Model onto a Relational Model" gave a brief discussion of a table layout for the tetrahedron-vertex adjacency relation (a de-normalized quadruple representation and its normalization.) SQL Server 2000 had indirect ways to switch between the normal and quadruple representations using a table-valued function or a triple join. As others [10] have pointed out, this translation can be easily expressed as a pivot operation. Using the `PIVOT` operator now supported in SQL 2005, the (de-normalized) quadruple representation of the tetrahedron-vertex relation can now be easily written as:

```sql
CREATE VIEW TetQuadRep AS
  SELECT ElemID, [0] AS v0, [1] AS v1, [2] AS v2, [3] AS v3
  FROM TetrahedronVertices
  PIVOT (SUM(VertexID) FOR Rank IN ([0], [1], [2], [3])) AS PVT
```

---

[2] FEA experts will have heard by now about "mesh-free" FE methods. Despite their limited success in solving non-academic problems, the underlying computational geometry problems turn out to be almost identical to the ones encountered in traditional FEA. Problem 3 (interpolation from a cloud of points) described below is a standard problem in the "mesh-free" arena.

Note that the `SUM` aggregate is completely artificial, since no real aggregation takes place. We could have used `MAX` or `MIN` as aggregates with the same result. This query, of course, performs MUCH better than a triple join described in Part I, and there is no longer a need to store both representations, as was necessary for large meshes in SQL Server 2000. Although the `UNPIVOT` operator can be used to switch from the quadruple representation to the normalized representation, the tetrahedron-vertex representation of Figure 1 should be the one stored, since the quadruple representation cannot be effectively indexed to allow for a fast tetrahedron lookup given a vertex.

**Getting Data into and out of the Database**

Given that you have a database schema defined, it is fairly easy to get data into or out of an SQL database. If you have an ASCII file named `C:\F.csv` containing fields separated by tab characters and records delimited by newline characters then the bulk copy command:

```
bcp DB.dbo.T in C:\F.csv –c –t \t –r \n -T
```

imports the data from file F to table T of database DB, and if file `F`'s data is binary then:

```
bcp DB.dbo.T in C:\F.dat -N -T
```

works faster by avoiding the ASCII to binary conversion costs (the `–c` means *character* the `–N` means *native* and the `–T` means *trusted*, use my credentials rather than requiring a password). Replacing "in" by "out" in the above commands exports the data in table T to file F. Within SQL one can issue a `BULK INSERT` command that works much like `bcp` but is inside the SQL process and so is more efficient.

Database table sizes are comparable to the binary file representation. In particular, a million row table of ten floating point columns occupies 80 MB in a binary file, 91 MB as an SQL table, and 192 MB in an ASCII file. Of course indices can increase the size of the SQL data, but those indices save IO.

Data loading is typically CPU bound. Our system can SQL `insert-select` a ten-column SQL table into another table at about 20 MB/s (250K records/second). Importing data from a binary file runs at half that speed (about 10 MB/s or 120 k records/s.) Importing data from an ASCII file uses nearly twice the CPU to do the conversion and so loads data at about 5MB/s (about 60k records/second). Running multiple import steps in parallel makes sense if you have two or more processors – since it is CPU bound, it gets nearly linear speedup.

The `bcp` tool and the `BULK INSERT` method do the trick for testing and development as well as the occasional bulk load. But, they are inadequate for building highly automated and robust integrated solutions. The actual workflow of Finite Element Analysis has the following data loading requirements:

- There are many files in many formats.

- The source files need to undergo certain transformations before they can be loaded.

- There is a precedence order in which the some files must be loaded and there are opportunities for concurrent loads (parallelism!).

- The source files may contain various errors and we need a reliable way to recover.

- Certain intermediate tasks need to be executed between loads.

- We don't want human file system monitors or event handlers.

This import-export process is repeated again and again as the model develops. Depending on the concurrency and error handling support of your favorite scripting language, just picture how much time and code you'll end up writing to produce a robust and fast (parallelism!) solution. These requirements are not unique to FEA, they re-appear in any computational science or analysis application, and they appear in any data warehouse application. So, as you might expect, database systems include a sophisticated tool suite to deal with the Extract-Transform-Load (ETL) problem. SQL Server has a component called *Integration Services* (SSIS) [11,16] that provides a graphical scripting, debugging, and monitoring system to design and execute ETL tasks. Figure 2 shows a typical SSIS package to move file data into the database.

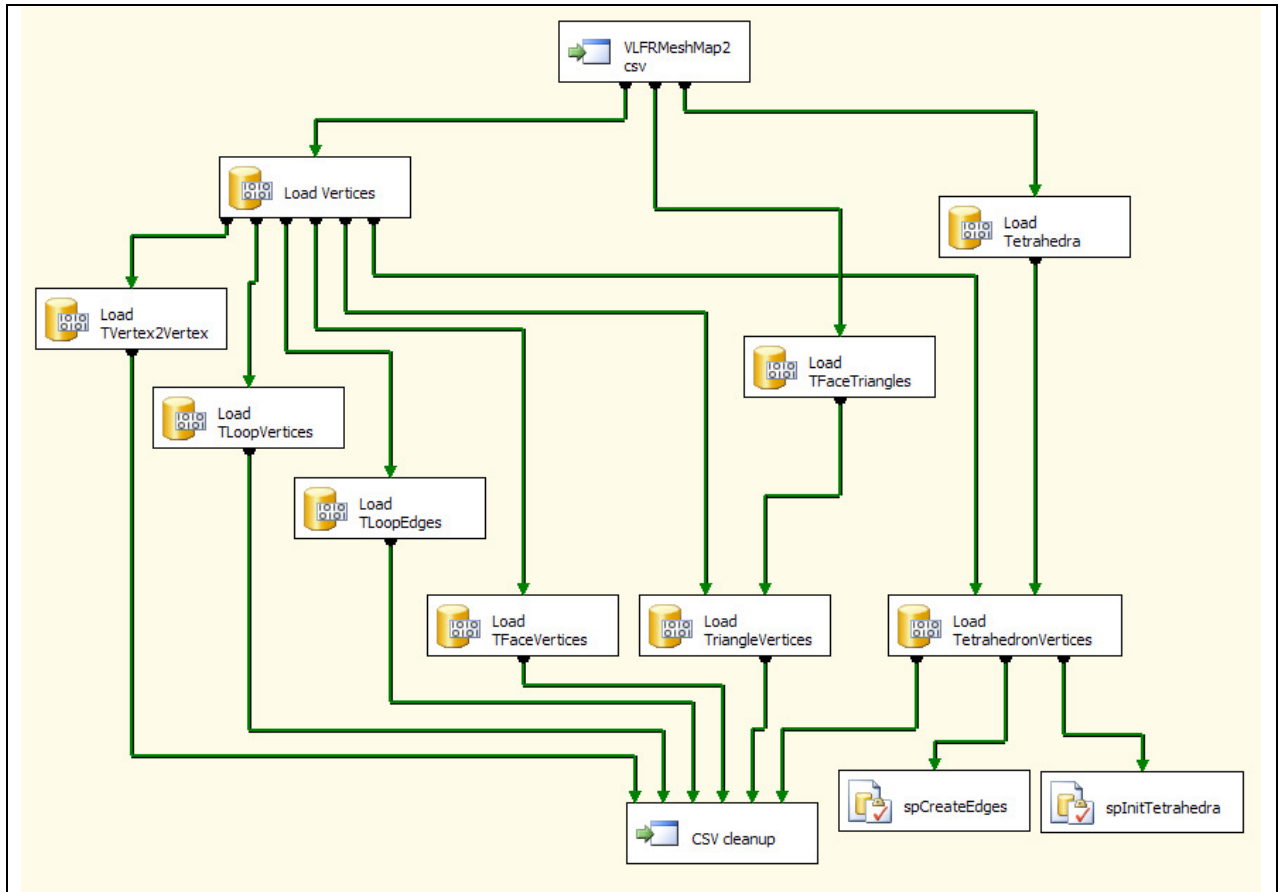

**Figure 2: A sample SSIS flow graph.** Each icon represents an SSIS task and the connections represent dependences. A task will not start unless all preceding tasks have completed successfully (failure dependencies are supported but not used in this package.) Independent tasks run in parallel. Most of the tasks shown are *Bulk Insert* tasks. There are also two *Execute Process* and two *Execute SQL* tasks. The first Execute Process task, "VLFRMeshMaps2csv", runs a Python script which splits and reformats the mesh generator output. All Bulk Insert tasks depend upon its successful completion. The second Execute Process task, "CSV cleanup", deletes intermediate processing files. It will run only if all Bulk Insert tasks have successfully completed. The two Execute SQL tasks execute stored procedures which initialize the mesh edges and tetrahedra. The SSIS package can be stored in a database or in a file. SSIS packages are typically executed from the command-line, the GUI, or a scheduled job under the control of the SQL Server Agent. SSIS packages are fully customizable through user-defined variables and transformations. For example, the package shown has 6 user-defined variables which parameterize the destination database server and database as well as the source directories for data, format files, python scripts and a filemask.

**Getting the Input Data to the Simulation**

Larger FEA typically execute as an MPI job on a Beowulf computer cluster. Traditionally, on simulation start, each node of the cluster reads its partition of the mesh from a specific file or from a specific offset within a file. The number of partitions may be less, equal, or greater than the number of MPI processes. The initial partitioning might have been obtained by a cheap method like recursive coordinate bisection [13] or it might be a left-over from a previous run, and the final partitioning is the result of a high-quality in-memory partitioner like ParMETIS [14]. Partitioning is easier and more flexible when the mesh is stored in a database indexed by some spatial parameter. We use a tetrahedron's center of mass (a point) on a Hilbert space-filling curve [5] as its spatial "key". Each tetrahedron's Hilbert code becomes the partitioning key. In this way it is easy to create any number of spatial partitions using the SQL Server 2005 row ranking functions (ROW_NUMBER, RANK, DENSE_RANK, NTILE). We use the NTILE as a simple partitioning routine. For example the query:

```
SELECT ElemID, NTILE(256) OVER (ORDER BY Hcode) AS partition FROM Tetrahedra
```

returns pairs of element ID and partition ID (between 1 and 256) based on the rank of a tet on the Hilbert curve. Partitioners based on space-filling curves are standard in parallel toolkits like Zoltan [13] and for most practical purposes the initial partitioning obtained this way has adequate balance. The primary goal here is to get a balanced startup scenario and nothing prevents us from adaptively re-partitioning in memory (with ParMETIS).

Partitioning is a good example of using the database to simplify the application design, execution and management. As explained before, many simulations spend considerable code and execution time on file handling. The use of a database system has the dual benefit of eliminating a lot of code and also giving automatic parallelism to reduce execution time.

Long-running simulations periodically create output (load steps or time steps) and check point/restart information. The compute node's local storage is insufficient to absorb the result sets from several days or weeks of processing which is why the results are written to dedicated I/O nodes or large scratch volumes. SQL Server Integration Services packages periodically wake up and copy recent simulation output data to the database where it becomes part of the model output and can be analyzed while the simulation progresses.

## How Databases Help Solve Some Standard Computational Geometry Problems in FEA[3]

Typical computational geometry problems in FEA deal with locating points inside elements and interpolating field values at (those) arbitrary points. Below, we formulate three typical FEA computational geometry problems and describe the implementation in SQL Server 2005 of Problem 1 in detail. We find that a set-oriented non-procedural language saves writing a lot of code, and overall executes much faster than a file-oriented approach. To be precise, the problem is not so much that traditional approaches store the data in files. The issues are the following:

1. In order to solve the problems described below, search and other auxiliary structures must be created in core. They must be created each time from scratch or they are persisted in some form adding more complexity. Building these indices takes programming time and execution time — in the worst case, every time a batch of objects is processed this execution overhead is incurred.

2. On a 32-bit architecture one quickly runs out of memory. Writing efficient out-of-core file processing code is not for the faint-hearted. Even on a 64-bit architecture, the sizes of datasets are often larger than the affordable RAM.

3. Parallelism can substantially increase throughput. Implementing such parallelism in a file-oriented design increases coding complexity.

Database systems create search structures (mostly indexes) as part of the initial setup. These structures work well even if most of the index and data are not memory-resident. SQL automatically allows parallelism with no extra programming if it is used as a set-oriented language. Unless record-at-a-time cursors are used in a SQL query, the optimizer automatically detects and exploits processor and disk parallelism.

Now let's look at some specific examples of using the database to simplify FEA programming.

---

[3] See Appendix B for a non-geometrical sample problem.

**Problem 1: The Point-in-Tetrahedron Query for Unstructured Tetrahedral Meshes**

Finding the mesh tetrahedron containing a given point is central to most FEA post-processing, especially visualization. Papadomanolakis et al. propose a novel solution to the point-in-cell problem for unstructured tetrahedral meshes called *Directed Local Search* (DLS) [7]. DLS bears little resemblance to traditional B-, R-, P-tree methods. It first makes a well-informed choice of a candidate tetrahedron and then uses a tetrahedron-connectivity-graph traversal technique to quickly find the containing tetrahedron. DLS is a hybrid algorithm exploiting both geometric and connectivity information. Papadomanolakis et al. implemented DLS in PostgreSQL [8] and compared it with other methods [7]. A detailed description of our DLS implementation in SQL Server 2005 is provided below.

**Problem 2: Interpolation of a Field Value at an Arbitrary Point from an Unstructured Mesh**

FEA approximates complex geometries and continuous fields by assigning values to isolated points in space (nodes). It uses interpolation to approximate the field's value at any point inside the mesh. In FEA terminology, the interpolation scheme is called a *shape function*. Roughly speaking, the value $f(p)$ of a field at a point $p$ inside an element $e$ is given as the weighted sum of the values $f_n$ attached to the nodes $N(e)$ of the element, like element corners and midpoints of edges.

$$f(p) = \sum_{n \in N(e)} s_n(p) f_n \qquad (1)$$

The weights $s_n(p)$ are the values of the element's shape functions $s_n$ at the point $p$. In particular, the shape functions are **not** constant across an element and, in general, are non-linear functions of a point's coordinates.

Field interpolation is an essential part of post-processing and visualization. Given an arbitrary point $p$, the point-in-tetrahedron query identifies the mesh element containing $p$. One can then compute the point's field values using equation (1) and the node's shape functions at $p$. If the shape functions are non-linear, a (small) system of non-linear equations needs to be solved to determine their values at $p$.

**Problem 3: Interpolation of a Value from a Cloud of Points**

Reference [12], formulates the value-from-a-cloud-of-points problem as follows: "Given a set of randomly distributed points $V = \{P_1, P_2, ..., P_n\}$ with associated scalar values $\{q_1, q_2, ..., q_n\}$ interpolate a value $q(x)$ at point $x$ within the convex hull of $V$." (The convex hull is the smallest convex set which contains the point set.) This problem naturally arises when experimental data must be interpolated onto a computer model. The model time-space geometry might be simplified and there is no one-to-one match between the measurement points and the model grid points.

Though formally very similar to the interpolation problem for unstructured meshes (Problem 2), the point cloud turns out to be more difficult. We compensate for the lack of pre-defined interpolation paths (connectivity) by introducing a spatial index (octree) to quickly find nearby points. The number of interpolation points depends on the desired interpolation order: Four points are adequate for or linear interpolation. More points are required for higher order interpolation. Unfortunately, proximity alone is not sufficient: certain dependences among the closest points like colinearity or complanarity will lead to singular covariance matrices in which case alternative points are required [12].

# The Point-in-Cell Query for Unstructured Tetrahedral Meshes using Directed Local Search (DLS)

The idea behind DLS can be stated as: let $p$ be a point with Cartesian coordinates $(x, y, z)$. Assume that we know *a priori* that there exists at least one mesh element that contains *p*. If the volume cannot be easily approximated by piecewise low-order or polygonal geometries, this decision is costly; but for our polycrystal models (see [4]) the answer is a trivial point-in-box test. The DLS method finds a tetrahedron containing $p$ in two steps:

1. Find a *good candidate* tetrahedron $t_c$ in the vicinity of $p$.
2. If $t_c$ turns out not to contain $p$, use the *tetrahedron-face connection graph* to pick an adjacent tetrahedron $t$ in the direction defined by the vector from the center of mass $c$ of the current tetrahedron to the point $p$ and repeat step 2 using $t$ as $t_c$ until the containing tetrahedron is found.

*Candidate Selection*

A common technique to find (small) spatial objects near a point $p$ is to compare their ranks on a Hilbert space-filling curve [5]. An (approximate) Hilbert curve can be used to define a ranking or linearization $H$ of points $(i, j, k)$ of a (non-negative) cubic lattice. The locality preserving nature of $H$ allows us to be optimistic: if the difference between the ranks $H(i, j, k)$ and $H(i_1, j_1, k_1)$ of two points is small, then their Euclidean distance between points $p = (i, j, k)$ and $p_1 = (i_1, j_1, k_1)$ is often small. This makes $H$ an excellent choice for a candidate selector. For each tetrahedron we pre-compute, store, and index (!) the $H$ code of its center of mass. If a point is close to the center of mass of a tetrahedron, then there is a fair chance (the bigger the tet the better!) that it falls inside the tetrahedron.

Note that neither a $H$-match nor an $H$-difference of a point and a tetrahedron's centroid yields any conclusion about the inclusion of the point in the tetrahedron. However, a simple algebraic test can establish that: Let $p_0, p_1, p_2, p_3$ be the corners of a tetrahedron and $e_1 = p_1 - p_0, e_2 = p_2 - p_0, e_3 = p_3 - p_0$ (vectors). If the tetrahedron is non-degenerate (= has non-zero volume), an arbitrary point $p$ can be uniquely written as

$$p = p_0 + \lambda e_1 + \mu e_2 + \nu e_3. \tag{2}$$

(This is a system of three linear equations for three unknowns $\lambda, \mu, \nu$.) $p$ is inside the tetrahedron defined by $p_0, p_1, p_2, p_3$, iff $\lambda, \mu, \nu \geq 0$ and $\lambda + \mu + \nu \leq 1$.[4] Solving for λ, μ, ν involves ~71 floating point computations (four $3 \times 3$ determinants and three divisions) – which are very fast on modern processors.

*Traversal*

If the candidate tetrahedron, $t_c$, contains the point $p$, then $t_c$ is the answer to the tet-contains-point query. If the containment test fails for $t_c$, the line from the centroid of $t_c$ to $p$ intersects at least one of the tetrahedron's faces.[5] Make the adjacent tetrahedron the new candidate. To implement this, let $c = \frac{1}{4}(p_0 + p_1 + p_2 + p_3)$ be the centroid of $t_c$. The ray defined by direction (vector) $p - c$ intersects the triangular facet $(p_i, p_j, p_k), 0 \leq i, j, k \leq 3$, iff $p$ can be represented as

$$p = c + \lambda p_i + \mu p_j + \nu p_k, \lambda, \mu, \nu \geq 0. \tag{3}$$

---

[4] An alternative way to establish the inclusion is to look at the signs of the (four) inner products of the inward pointing normals of a tetrahedrons triangular faces with the vectors from the face centroids to $p$. $p$ is inside, iff all of them are non-negative.

[5] The ray might intersect an edge or a vertex in which case there are multiple choices.

We examine each of the four faces of $t_c$ until we find a solution to this equation. Unless the triangle we find is on the outer surface of the body, there is exactly one tetrahedron on the other side of the intersected face.[6] We make this neighbor our current tetrahedron and repeat the inclusion test. Figure 3 illustrates the traversal technique in 2D.

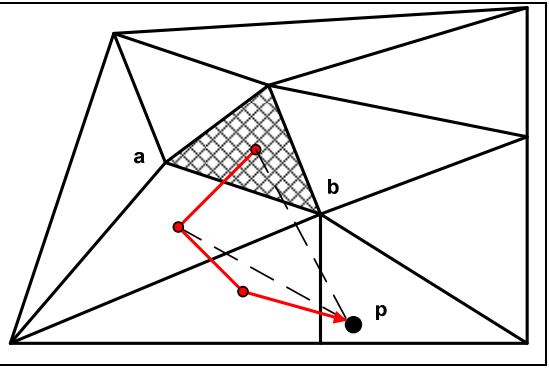

**Figure 3: A sample 2D traversal.** The shaded triangle represents the candidate triangle. It does not contain $p$. The line from the center of gravity of the candidate to $p$ intersects the edge $\overline{ab}$. The triangle on the other side of $\overline{ab}$ does not contain $p$ either and we have to continue iterating until we reach the triangle that contains $p$.

**A DLS Implementation in SQL Server 2005**

Based on the description in the previous section, the following routines need to be implemented:

- A function to calculate the Hilbert code, $H$, of a point $(x, y, z)$
- A Boolean function that implements the point-in-tet test
- A routine that, given a ray *r* and a tetrahedron *t*, determines which facet(s) of *t* are intersected by *r*.
- A routine which determines the ID of the tetrahedron on the other side of a tetrahedron face

*Calculating the $H$ code*

Reference [5] describes an elegant implementation of $H$. We converted it to C# (about 500 lines) and made some minor adaptations: In the C implementation of [5], the $H$ code is represented as a structure of three unsigned 32-bit integers. The largest scalar integral type in SQL is a signed 64-bit `bigint`. We condensed three unsigned 20-bit integers into a `bigint` $H$ code. This gives us a resolution of $2^{21} - 1 = 2,097,151$ grid points in each coordinate direction. The C# prototype looks as follows:

```
public class Hilbert3D {
  // we use only the lower 20 bits of x, y, z
  public static UInt64 H_encode(UInt32 x, UInt32 y, UInt32 z);
  // this is what we will call from T-SQL
  [SqlFunction(IsDeterministic = true, IsPrecise = true, DataAccess = DataAccessKind.None)]
  public static SqlInt64 H_encode_SQL(SqlInt32 x, SqlInt32 y, SqlInt32 z) {
    return H_encode((UInt32)x, (UInt32)y, (UInt32)z); }
}
```

This defines the `H_encodeSQL` function signature and tells SQL that the function is *deterministic* and *precise*, so that it can optimize the function call and execute it in parallel.

After registering the DLL with SQL Server, the encoding function can be made accessible to T-SQL via:

```
CREATE FUNCTION fnH_encode(@x int, @y int, @z int)
RETURNS bigint AS EXTERNAL NAME [SFC].[SFC.Hilbert3D].[H_encode_SQL]
```

(The DLL's name is `SFC.dll` and `Hilbert3D` is a class in a namespace called `SFC`.)

---

[6] For simplicity, cavities, holes and the like are ignored here. The algorithm can be easily modified to handle those cases [7].

We calculate $H$ for each tetrahedron and create a clustered index on the `Hcode` column:

```sql
CREATE TABLE Tetrahedra (
  ElemID int    PRIMARY NONCLUSTERED KEY,
  X      float  NOT NULL,
  y      float  NOT NULL,
  z      float  NOT NULL,
  Hcode  bigint NOT NULL,
...)
UPDATE Tetrahedra SET Hcode = dbo.fnH_encode(x, y, z)
CREATE CLUSTERED INDEX Idx_Hcode ON Tetrahedra(Hcode)
```

The $H$ calculation takes about $22\mu s$ per row. An update takes about $297\mu s$ per row.[7] As pointed out in [7], a *partial* index would be sufficient. For example, randomly selecting an adjacent tetrahedron for each vertex in the mesh would provide an adequate subset to index and to make DLS work.

*Point-in-Tet Test*

For the point-in-tet test we solve equation (2) and check whether $\lambda, \mu, \nu \geq 0$ and $\lambda + \mu + \nu \leq 1$. The test returns a T-SQL Boolean value indicating the result. Note that the test also returns 1 also if the point falls on the tetrahedron's face, edge, or vertex. To accomplish that (in particular for high aspect ratio tets!), the inequalities for $\lambda, \mu, \nu$ are augmented with a tolerance $\varepsilon$. A value of $\varepsilon = 10^{-15}$ worked well for us in practise. The C# prototype of this T-SQL user defined function is:

```sql
CREATE FUNCTION fnPointInTet(@ElemID int, @x float, @y float, @z float) RETURNS bit
AS EXTERNAL NAME FEMUtils.FEMUtils.Tetrahedron.PointInTet
```

reads as follows:
```csharp
[SqlFunction(IsDeterministic = true, IsPrecise = true, DataAccess = DataAccessKind.Read)]
public static SqlBoolean PointInTet(SqlInt32 ElemID, SqlDouble x, SqlDouble y, SqlDouble z);
```

Note that the data access attribute is now set to `Read` since inside the method we have to fetch the corner coordinates in order to formulate equation (2). Since the code is executing on the server in the context of the user-defined funtion, there is no need to create a special loopback connection. We bind to this context connection as follows:

```csharp
using (SqlConnection conn = new SqlConnection("context connection=true;"))
```

*The RayIntersectsTetFace Test*

We determine which face is intersected by solving equation (3) for each triangular face until we find one where $\lambda, \mu, \nu \geq 0$. The test returns the local face ID (0,1,2,3) of the intersected face or returns 4, if the point is actually inside the tetrahedron:

```sql
CREATE FUNCTION fnRayIntersectsTetFace(@ElemID int, @x float, @y float, @z float) RETURNS tinyint
AS EXTERNAL NAME FEMUtils.FEMUtils.Tetrahedron.RayIntersectsTetFace
```

The C# prototype is virtually identical to the point-in-tet test, except that the return value is of type `SqlByte` to represent a `tinyint` in T-SQL.

```csharp
[SqlFunction(IsDeterministic = true, IsPrecise = true, DataAccess = DataAccessKind.Read)]
public static SqlByte
            RayIntersectsTetFace(SqlInt32 ElemID, SqlDouble x, SqlDouble y, SqlDouble z);
```

*The GetTetFaceNeighbor Lookup*

Papadomonolakis [7] suggests storing the IDs of each tetrahedron's face-neighbors. We do not believe that this is practical: mesh generators typically do not generate this information and calculating it from scratch for large meshes is expensive – especially since little of the information would ever be used. A lazy evaluation strategy, one where we compute the neighbors as needed and save them for future lookup, seems more reasonable.

---

[7] The PostgreSQL implementation described in [7] does not store the Hilbert codes: the B-tree is built on Hilbert order and an appropriate comparison function is used in the B-tree traversal. SQL Server does not give us this level of control and we have to mimic similar behavior through a clustered index.

We wrote a T-SQL function `fnGetTetFaceNeighbor` which, given an element ID and the rank of a face, determines the ID of the tetrahedron on the other side or returns -1 in case the face is a triangle on the surface. At its heart is the `TetrahedronVertices` relation:

```sql
CREATE TABLE TetrahedronVertices (
  ElemID      int       NOT NULL REFERENCES Tetrahedra,
  Rank        tinyint NOT NULL CHECK(Rank BETWEEN 0 AND 3),
  VertexID    int       NOT NULL REFERENCES Vertices,
  CONSTRAINT PK_TetrahedronVertices PRIMARY KEY (ElemID, Rank),
  CONSTRAINT UNQ_TetrahedronVertices UNIQUE (ElemID, VertexID)
)
```

The following index improves the performance of `fnGetTetFaceNeighbor` by giving a fast adjacent element lookup given a vertex:

```sql
CREATE INDEX Idx_VertexID ON TetrahedronVertices (VertexID)
```

The $i$-th face of a tetrahedron $(i = 0,1,2,3)$ is the face opposite to the $i$-th corner. Given an element and one of its vertices, we are looking for another element which shares the other three vertices. Unless the face opposite to the vertex is on the outer surface, there is exactly one tetrahedron with that property, and its ID is what the `fnGetTetFaceNeighbor` T-SQL function below returns:

```sql
CREATE FUNCTION fnGetTetFaceNeighbor(@ElemID int, @oppositeVertex tinyint) RETURNS int AS
BEGIN
  DECLARE @result int           -- result is ID of desired tet
  SELECT @result = ElemID        -- find the first face
  FROM TetrahedronVertices       -- in the Tet-Vertex table
  WHERE VertexID IN (            -- where all the vertices in common with
    SELECT VertexID              -- one of the vertices of @ElemID
    FROM TetrahedronVertices     -- (this select statment returns
    WHERE ElemID   = @ElemID     -- all the vertices of the tet)
      AND Rank    != @ oppositeVertex) -- excluding the one opposite the face
  AND ElemID != @ElemID          -- it is a different face
  GROUP BY ElemID                -- group matching vertices
  HAVING COUNT(VertexID) = 3     -- insist all 3 verticies match
  RETURN COALESCE(@result, -1)   -- return -1 if no such element
END
```

The $i$-th face of a tetrahedron $(i = 0,1,2,3)$ is the face opposite to the $i$-th corner. Given an element and one of its vertices, we are looking for another element which shares the other three vertices. Unless the face opposite to the vertex is on the outer surface, there is exactly one tetrahedron with that property, and its ID is what `fnGetTetFaceNeighbor` returns.

**Performance of DLS**

One of the metrics suggested in [7] is *page accesses per point query*. Since this report is aimed at end-users and we assume that our metadata (the mesh) will often fit in memory, we prefer a more application-oriented metric, namely *points per second*. Our test setup is as follows: We consider a fixed mesh from one of our polycrystal models (see Figure 4).

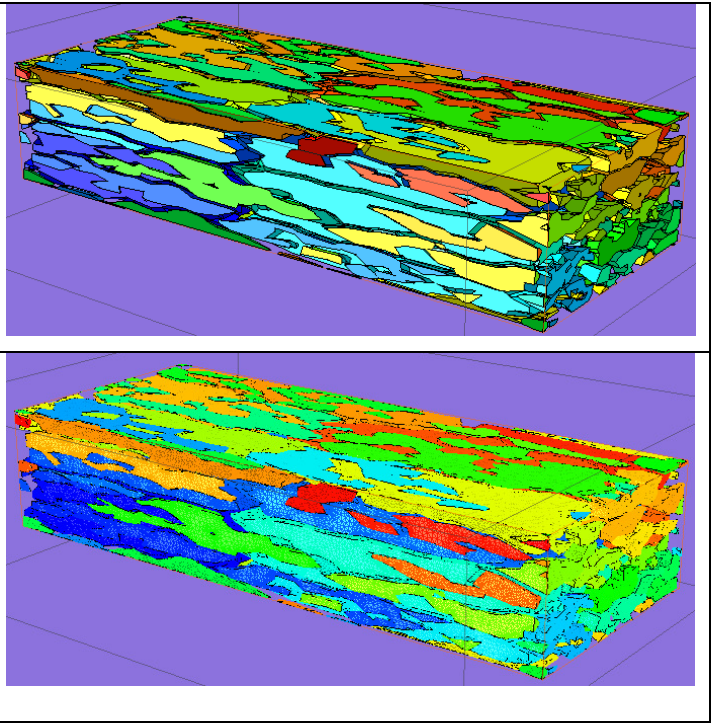

**Figure 4: Sample model and mesh.** In the top figure, a polycrystal geometry [4] is shown. (Data by courtesy of A.D. Rollett and S. Sintay, Carnegie Mellon University) The individual crystals or grains have been isolated for visualization purposes. The figure below shows the surface mesh. The grains themselves are filled with tetrahedra, abutting the triangles on the surface. The mesh has 8,782,315 tetrahedra and 1,520,308 vertices. (In Appendix B, we describe a procedure how to generate the triangle set on the outer surface of the polycrystal with a simple SQL query.) Because of the rich surface structure of the grains, the triangle (and tetrahedron) density is highly non-uniform. The non-uniformity in object density and the unbounded discrepancy in shape between a coordinate-plane aligned box and a tetrahedron is what makes the point-in-cell test for unstructured tetrahedral meshes so difficult.

The mesh shown in Figure 4 has 8,782,315 tetrahedra and 1,520,308 vertices. It is very non-uniform in terms of element size and connectivity. A vertex is shared on average by 23 tetrahedra, with a range from 2 to 192 and a variance of 67. The geometry of this particular grid is a brick with the dimensions: $[0,5] \times [-0.95, 2.783] \times [-0.283, 1.258]$.

For each pair of center $c = (cx, cy, cz)$ and radius $r$, we generated a table `T(x float, y float, z float)` containing a spherical cloud of uniformly distributed random points. That is, the points $(x, y, z)$ in the cloud satisfy:

$$(x-cx)^2 + (y-cy)^2 + (z-cz)^2 \leq r^2 \qquad (4)$$

Each table contained 20,000 points and for each such table `T` we ran the following query:

```
SELECT COUNT(DISTINCT dbo.fnGetTet4Point(x, y, z)) FROM T
```

(`fnGetTet4Point` is the user-defined function that wraps all of the above functions and returns the ID of a tetrahedron containing the point $x, y, z$.) This query returns the number of distinct tetrahedron IDs encountered when categorizing the points of the point cloud from the sample table.

Table 1 shows the elapsed runtime and points per second (execution time divided by the number of points in the table) for different centers and radii. We also recorded the average number of number of distinct tetrahedron IDs returned by the query.

| Table 1: Points per second and distinct tet count for sample clouds. | | | |
|:---:|:---:|:---:|:---:|
| $c$ | $r$ | Points / s | #Distinct Tetrahedra |
| (2.1, 1.35, 0.65) | 0.00001 | 2857 | 1 |
| (2.1, 1.35, 0.65) | 0.0001 | 2222 | 2 |
| (2.1, 1.35, 0.65) | 0.001 | 1818 | 2 |
| (2.1, 1.35, 0.65) | 0.01 | 741 | 8 |
| (2.1, 1.35, 0.65) | 0.1 | 689 | 294 |
| (2.1, 1.35, 0.65) | 0.25 | 689 | 1831 |
| (2.1, 1.35, 0.65) | 0.5 | 476 | 2443 |
| (2.1, 1.35, 0.65) | 0.6 | 426 | 2889 |
| (0, 0, 0) | 0.5 | 416 | 3249 |

The results in Table 1 confirm what one would expect: the smaller the radius and the fewer distinct tetrahedra the higher the throughput (points per second). In that sense, the first result in Table 1 comes close to "peak performance" where we are dealing effectively with one tetrahedron. We expect the cache hit rate to degrade with increasing radius and an increasing number of distinct tetrahedra in the result set. The exact degradation is controlled by the local geometry, i.e., the number, size, and shape of the tetrahedra near $c$ and the random points.

To account for the local variability in geometry we generated point clouds of random diameter in random locations. The maximum radius depends on the model scale and the application. For our model scale and the intended application (low order screening) a maximum radius of 0.1 is more than generous. Given a fixed total number of points (20,000), the locality behavior will change depending on whether we have more points clustered in fewer locations or fewer points clustered in more locations. Table 2 shows the results for randomly centered and sized point clouds: we generated $N$ uniformly distributed points $c$ and radii $r$ (with mean $\mu(r)$ and standard deviation $\sigma(r)$), and $N_{c,r}$ points per cloud, where $N \cdot N_{c,r} = 20,000$.

| Table 2: Points per second and distinct tet count for random sample clouds | | | | | |
|:---:|:---:|:---:|:---:|:---:|:---:|
| $N$ | $N_{c,r}$ | $\mu(r)$ | $\sigma(r)$ | Points / s | #Distinct Tetrahedra |
| 10 | 2000 | 0.0312 | 0.0130 | 833 | 54 |
| 20 | 1000 | 0.0299 | 0.0142 | 588 | 159 |
| 200 | 100 | 0.0267 | 0.0148 | 250 | 1398 |
| 2000 | 10 | 0.0246 | 0.0145 | 208 | 4389 |
| 10 | 2000 | 0.0623 | 0.0261 | 556 | 138 |
| 20 | 1000 | 0.0598 | 0.0284 | 571 | 412 |
| 200 | 100 | 0.0534 | 0.0296 | 312 | 2412 |
| 2000 | 10 | 0.0491 | 0.0289 | 167 | 5405 |
| 20000 | 1 | N/A | N/A | 75 | 7392 |

The `Hcode` (clustered) index was very selective in these experiments: no two tetrahedra had the same `Hcode`.

### Practical Considerations

Officially only assemblies from source code written in C# and/or VB.NET are supported in SQL Server 2005. Our attempts to deploy assemblies generated with /clr:safe from C++/CLI [17] codes failed repeatedly. For scientific and engineering applications, this is a severe limitation: there is a substantial C++ code base and the conversion from C++ templates to C# generics is tedious and, in some cases, impossible.

When creating an assembly in SQL Server, access security permissions must be specified. The options are SAFE, EXTERNAL ACCESS, and UNSAFE. SAFE is the default. All the examples in this report are SAFE, no access to the file system, the network, or unmanaged code was allowed or requried. EXTERNAL ACCESS relaxes this restriction and allows access, for example, to the file system through managed and verifiable .NET interfaces. Finally, the UNSAFE permission set allows calling unmanaged code (e.g. using P/Invoke). This creates several opportunities to integrate data from diverse sources under the umbrella of SQL Server.[8] For example, data can be read from or written to netCDF [20] files by calling routines in the (unmanaged) netCDF DLL from table-valued functions or stored procedures.

Got XML data? There is no need anymore to work with textual representations of XML documents or shred them to fit a tabular model. It is beyond the scope of this report, but we don't want to close without encouraging you to explore SQL Server's XML type and XQuery support (and many of the other new features!).

### Summary

This discussion of the database aspects of FEA started by explaining that many current programs spend considerable code and time on data parsing and that the evolution of these programs is hampered by need to make bi-lateral changes to the data producer and the data consumer. Using self-defining data schemas and a relational database interface gives FEA codes the same data independence advantages that they give to other applications. These data independence and schema evolution issues are particularly important for the FEA mesh and its associated attributes.

Databases tools are also quite useful for data ingest. The next article in this series shows how helpful the database representation is for analysis and visualization tools. But, for the simulation steps, the most important aspects are the data independence, parallel database access (for performance), and the ease with which one can specify fairly complex data access concepts. This article gave several examples of complex data access — ranging from the dual representation of tetrahedra using the pivot operator – the normalized face representation and the denormalized quad representation. It also showed how directed local search leverages relational operators and indices to quickly implement spatial search.

These are not isolated examples, they are chosen as representatives form a large collection of techniques we have used to build the FEA system behind our Computational Materials system COMPASS [25]. That system is a considerable advance over its predecessor and demonstrates the benefits of using a database representation for the FEA model, grid, results, *and beyond*.

### Acknowledgements

We gratefully acknowledge the support of the Cornell Theory Center, Microsoft, and Microsoft Research. The first author would like to acknowledge the support received under the DARPA SIPS program.

---

[8] We also have not discussed SQL Server's ability to host XML Web service endpoints without IIS.

## Appendix A: System Configuration

The system used for all the tests in this report in its configuration and performance is very similar to the one dubbed "Tyan/Opteron/Marvell Configuration" in [2].

**Hardware**

- Tyan Thunder K8SD Pro S2882-D motherboard
- 2 x Dual Core AMD Opteron 275 (2.2 GHz, 1 MB L2 cache per core)
- 8 x 1 GB DDR-SDRAM memory
- 3 x Supermicro 8-port SATA card (AOC-SAT2-MV8)
- 24 x Hitachi 500GB HDS725050KLA360 hard drives

**Software**

- Microsoft Windows Server 2003 R2 Enterprise x64 Edition
- Microsoft SQL Server 2005 Enterprise x64 Edition
- Microsoft .NET Framework 2.0 (build 2.0.50727)

The attached disk storage is configured as JBOD (24 NTFS mount points). We ran SQLIO [24] to determine how the aggregate bandwidth scales across the three SATA controllers (see Figure 5).

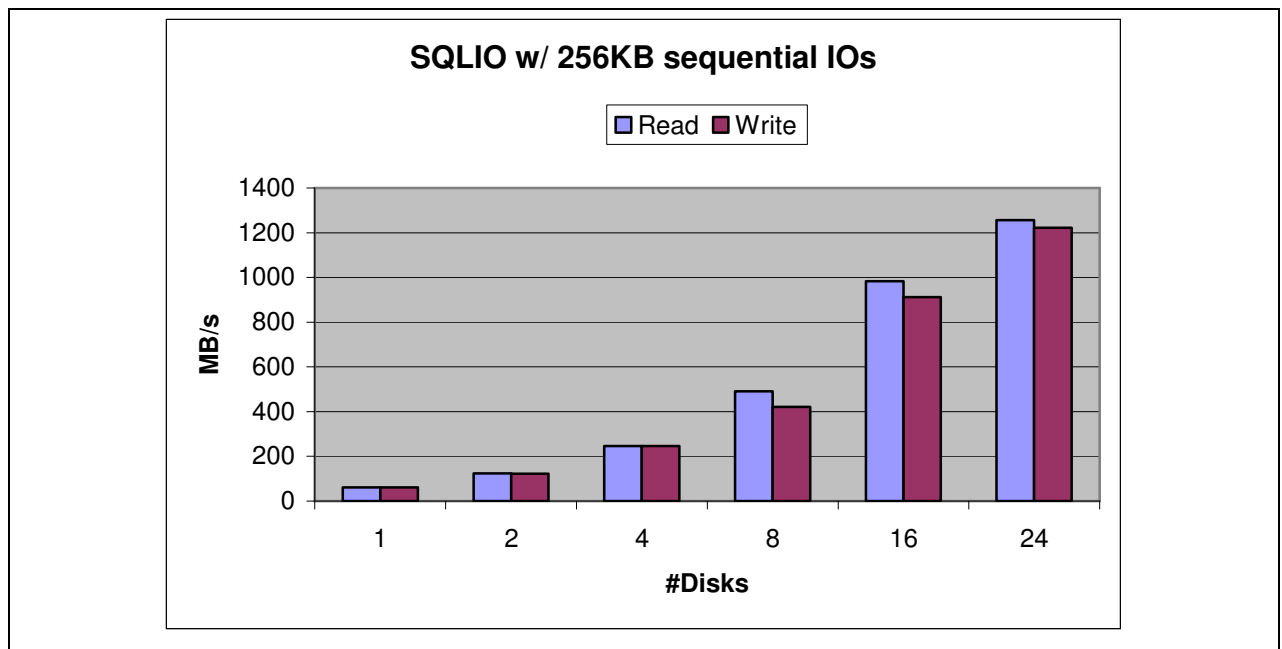

**Figure 5: SQLIO performance.** The bandwidth was determined by reading or writing for 30 seconds 256KB blocks sequentially to files on multiple disks. Both read and write aggregate bandwidth scale almost linearly across all three controllers and 24 disks. The single disk performance for read and write was slightly over 61 MB/s. The top rate across 24 disks was around 1.2 GB/s for read and write. The CPU utilization while simultaneously accessing all 24 disks was around 18% (4 cores = 100%).

## Appendix B: Recovering the Surface Mesh of a Tetrahedral Mesh

A common (non-geometric) problem in FEA is recovering the surface mesh given only a tetrahedral mesh. Each tetrahedron is bounded by four triangular facets and such a facet is considered part of the surface mesh, if and only if it bounds exactly one tetrahedron. (Internal triangular facets are shared by exactly two tetrahedra: a triangular facet shared by less than one or more than two tetrahedra indicates an error in the connectivity.) A standard approach is to create a hash table where the keys are triangles and the values count how many elements share the key triangle. If the connectivity is correct, there are only two possible values (1 and 2) and, after all triangles have been processed, the triangle keys with value 1 are the ones that make up the surface mesh. There are a few subtleties and refinements of the problem that complicate matters somewhat:

1. A triangle is typically represented as a triplet of its corner vertex IDs $(a,b,c)$. Do $(a,b,c)$ and $(b,a,c)$ represent the same triangle? It depends…

2. The surface of a volume mesh is typically *oriented*, i.e., for each triangle on the surface we pick an order of its vertex IDs such that all resulting geometric normals either point into the interior of the (volumetric) domain or into the exterior.[9] Under these assumptions, $(a,b,c)$ and $(a,c,b)$ do not represent the same (oriented) triangle, because the normal (vector product) $\vec{ab} \times \vec{ac}$ has the opposite sign of $\vec{ac} \times \vec{ab} = -\vec{ab} \times \vec{ac}$. Incoherently oriented surface meshes may cause rendering (shading) artifacts in visualization applications and incorrect results in numerical integration — we typically have a keen interest in generating a coherently oriented surface mesh.

We solve this problem in two steps: we first generate a set of normalized un-oriented triangles representing the (un-oriented) surface mesh and then, in a second step, recover the orientations from the tetrahedral volume mesh.

The triplet representation $(a,b,c)$ can be easily normalized[10] by imposing a constraint like $a < b < c$. The set of (normalized) triplets corresponding to the triangles in the surface mesh can be easily generated as follows:

```sql
SELECT A.VertexID AS [0], B.VertexID AS [1], C.VertexID AS [2] -- select the 3 verticies
FROM TetrahedronVertices AS A                                  -- by insisting they are part of
  JOIN TetrahedronVertices AS B ON A.ElemID = B.ElemID          -- the same element
  JOIN TetrahedronVertices AS C ON B.ElemID = C.ElemID          -- and
WHERE A.VertexID < B.VertexID AND B.VertexID < C.VertexID      -- that thery are in ascending order
GROUP BY A.VertexID, B.VertexID, C.VertexID                    -- consider all triplets
HAVING COUNT(A.ElemID) = 1                                     -- that occurr in a single element
```

The optimizer creates a rather efficient execution plan for this query: for our volume mesh of 8,782,315 tetrahedra, 147,708 triangles are generated in 290 seconds at an average CPU utilization above 97% (each CPU core accounts for 25%). If all we want is an un-oriented surface mesh, we are done. To get the orientations right, we have to understand where they come from and why the triplet representation is a little off the mark.

A tetrahedron can be represented as a quadruple of its corner vertex IDs $(a,b,c,d)$. Strictly speaking, an *oriented* tetrahedron is an equivalence class of such quadruples: two such quadruples are considered equivalent if they differ only by an even permutation[11] of their entries (components). For example, the quadruples $(0,1,2,3)$ and $(1,2,0,3)$ represent the same oriented tetrahedron. On the other hand, the quadruple $(1,0,2,3)$ represents a tetrahedron of the opposite orientation (since a transposition is an odd permutation). The oriented boundary $\partial(a,b,c,d)$ of an oriented tetrahedron $(a,b,c,d)$ consists of four oriented triangles and can be symbolically written as (see Figure 6):

---

[9] We ask the mathematically inclined reader to bear with us: the suggested correlation between orientability and two-sidedness of a closed surface is an artifact of the three-dimensional Euclidean space. Orientability is an "inner" topological property, single- or two-sidedness is not. All of the following do exist: orientable two-sided faces, non-orientable single-sided faces, orientable single-sided faces, and non-orientable two-sided faces. [22,23]

[10] We do not mean *normalized* in the database sense of normal forms, but *unique*.

[11] The swap of two successive entries in an n-tuple is called a *transposition*. For example, (1,3,2,4) is a transposition of (1,2,3,4). It can be shown that an arbitrary permutation of the entries can be represented as a sequence of transpositions. [21] A permutation is called *even*, iff it can be represented as an even number of transpositions. Otherwise it is called *odd*. For example, (2,3,1) is an even permutation of (1,2,3), whereas (1,3,2) is an odd permutation of (1,2,3).

$$\partial(a,b,c,d) \equiv (b,c,d)-(a,c,d)+(a,b,d)-(a,b,c) \qquad (5)$$

The minus signs in this expression indicate orientation reversal: for example, the triangle $(c,a,d)$ is on the boundary of the (oriented) tetrahedron $(a,b,c,d)$ and **not** $(a,c,d)$.

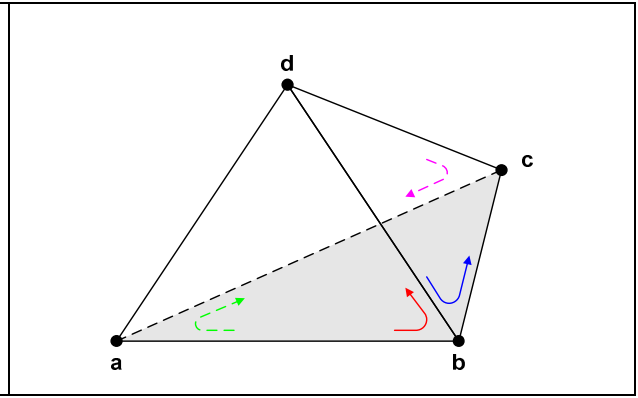

**Figure 6: Orientation of tetrahedron facets.** The orientation of the triangular facets of a tetrahedron according to equation (5) is shown. The order of the vertex IDs in the triplet can be interpreted as the order in which to "visit" the triangle corners. For example, the red arrow indicates the traversal direction for triangle $(a,b,d)$. Since the starting point of the traversal is arbitrary, $(b,d,a)$ and $(d,a,b)$ represent the same oriented triangle (even permutations). $(a,d,b)$, on the other hand, represents the same triangle with the opposite orientation.

According to equation (5), the triangles on the boundary of an oriented tetrahedron inherit their orientation from the tetrahedron. To correctly orient our surface triangles, we must go back to the adjacent tetrahedra and check their orientation. Getting back to the tetrahedra from the set of normalized, un-oriented triplets is a little less straightforward than we would like: no matter which representation of the tetrahedra we use, `TetrahedronVertices` or `TetQuadRep`, we cannot link them up directly to sorted triplets. At this point, SQL's `UNPIVOT` operator comes in handy to (DB-) normalize our triangle set:

```sql
SELECT TriID, VertexID
  INTO #triangles
  FROM (
    SELECT ROW_NUMBER() OVER (ORDER BY [0], [1], [2]) AS TriID, [0], [1], [2]
    FROM (
      SELECT A.VertexID AS [0], B.VertexID AS [1], C.VertexID AS [2]
      FROM TetrahedronVertices AS A
        JOIN TetrahedronVertices AS B ON A.ElemID = B.ElemID
        JOIN TetrahedronVertices AS C ON B.ElemID = C.ElemID
      WHERE A.VertexID < B.VertexID AND B.VertexID < C.VertexID
      GROUP BY A.VertexID, B.VertexID, C.VertexID
      HAVING COUNT(A.ElemID) = 1) AS D
  ) AS E
  UNPIVOT (VertexID FOR Rank IN ([0], [1], [2])) AS unpvt
CREATE INDEX IDX_VertexID ON #triangles(VertexID)
```

All the `UNPIVOT` operator does is to split each triplet into three rows, one for each vertex. We also used the `ROW_NUMBER` ranking function to create a temporary triangle identifier. Since we will reuse this set a few times, it is worthwhile creating a temporary table and creating an index on the `VertexID` column (see below).

Joining the `#triangles` table with the `TetrahedronVertices` relation gets the orientations right; however, there is one final twist. Different visualization or FEA packages often have different conventions of how to number or rank entities. A tetrahedron has on its boundary four corners, six edges, and four triangles: within each category, which is first, second etc.? This is clearly arbitrary but necessary to make any practical implementation work. We follow the convention used in P. Wawrzynek's FemLib [18] library. Let $(v_0, v_1, v_2, v_3)$ be a tetrahedron. It is FemLib's convention that the four oriented triangles on its boundary are given (in order) by: $[(v_0, v_1, v_2), (v_1, v_3, v_2), (v_2, v_3, v_0), (v_0, v_3, v_1)]$. For example, the third facet of tetrahedron (12, 4711, 841, 3) would be (841, 3, 12). Since both tetrahedra and triangles are stored in (DB-)normal form, we also need a mapping between tetrahedron local vertex IDs (which range from 0 to 3) and triangle local vertex IDs (which range from 0 to 2). We store the FemLib convention and this mapping in a small lookup table as follows:

```sql
CREATE TABLE FemLibTetFaces (
  TetFaceRank   tinyint NOT NULL CHECK(TetFaceRank BETWEEN 0 AND 3),
  TetVertexRank tinyint NOT NULL CHECK(TetVertexRank BETWEEN 0 AND 3),
  TriVertexRank tinyint NOT NULL CHECK(TriVertexRank BETWEEN 0 AND 2),
```

```
    CONSTRAINT PK_FemLibTetFaces PRIMARY KEY (TetFaceRank, TetVertexRank)
)
```

The table has only twelve rows (three entries for each of the four faces):

| TetFaceRank | TetVertexRank | TriVertexRank |
|---|---|---|
| 0 | 0 | 0 |
| 0 | 1 | 1 |
| 0 | 2 | 2 |
| 1 | 1 | 0 |
| 1 | 3 | 1 |
| 1 | 2 | 2 |
| 2 | 2 | 0 |
| 2 | 3 | 1 |
| 2 | 0 | 2 |
| 3 | 0 | 0 |
| 3 | 3 | 1 |
| 3 | 1 | 2 |

The final query looks as follows:

```
SELECT C.ElemID, TriVertexRank AS Rank, VertexID
FROM (
  SELECT ElemID, 0 AS TetFaceRank
  FROM #triangles AS A JOIN TetrahedronVertices AS B ON A.VertexID = B.VertexID
  WHERE Rank != 3 GROUP BY ElemID, TriID HAVING COUNT(A.VertexID) = 3
  UNION ALL
  SELECT ElemID, 1 AS TetFaceRank
  FROM #triangles AS A JOIN TetrahedronVertices AS B ON A.VertexID = B.VertexID
  WHERE Rank != 0 GROUP BY ElemID, TriID HAVING COUNT(A.VertexID) = 3
  UNION ALL
  SELECT ElemID, 2 AS TetFaceRank
  FROM #triangles AS A JOIN TetrahedronVertices AS B ON A.VertexID = B.VertexID
  WHERE Rank != 1 GROUP BY ElemID, TriID HAVING COUNT(A.VertexID) = 3
  UNION ALL
  SELECT ElemID, 3 AS TetFaceRank
  FROM #triangles AS A JOIN TetrahedronVertices AS B ON A.VertexID = B.VertexID
  WHERE Rank != 2 GROUP BY ElemID, TriID HAVING COUNT(A.VertexID) = 3
) AS C
    JOIN FemLibTetFaces AS D ON C.TetFaceRank = D.TetFaceRank
    JOIN TetrahedronVertices AS E ON C.ElemID = E.ElemID AND D.TetVertexRank = E.Rank
```

On our test machine, for the given mesh, the query runs in about 18 seconds, which brings the total time for the (oriented) surface recovery to around 310 seconds or a little over 5 minutes.[12]

Clearly, this solution can be easily generalized for meshes with mixed element types (tets, bricks, wedges, and pyramids) and triangles and quadrilaterals in the surface mesh. For a modest amount of SQL, we get a fast and highly parallel solution!

---

[12] T-SQL in SQL Server 2005 supports so-called *Common Table Expressions* (CTE). Though tempting, using a CTE instead of the temporary table is not a good idea (in this example): the index on the vertex ID column of the table is crucial for performance.